\documentstyle[aps,preprint]{revtex}
\begin{document}
\author{R. Allub\thanks{%
Member of the Carrera del Investigador Cient\'{\i}fico del Consejo Nacional
de Investigaciones Cient\'{\i}ficas y t\'{e}cnicas (CONICET).} and B. Alascio%
$^{*}$}
\title{Effect of double exchange and diagonal disorder on the magnetic and
transport properties of La$_{1-x}$Sr$_x$MnO$_3$.}
\address{Centro At\'{o}mico Bariloche, (8400) S. C. de Bariloche, Argentina.}
\maketitle

\begin{abstract}
We use a model previously formulated based on the double exchange mechanism
and diagonal disorder to calculate magnetization and conductivity for La$%
_{1-x}$Sr$_x$MnO$_3$ type crystals as a function of temperature. The model
represents each Mn$^{4+}$ ion by a spin S=1/2, on which an electron can be
added to produce Mn$^{3+}$. We include a hopping energy $t,$ two strong
intratomic interactions: exchange $J$, and Coulomb $U,$ and, to represent in
a simple way the effects of disorder, a Lorentzian distribution of diagonal
energies of width $\Gamma $ at the Mn sites.

In the strong coupling limit, $J,U>>t,\Gamma $, the model results can be
expressed in terms of $t$ and $\Gamma .$

We use the results of the model to draw ''phase diagrams'' that separate
ferromagnetic from paramagnetic states and also ''insulating '' states where
the Fermi level falls in a region of localized states from ''metallic ''
where the Fermi level falls in a region of extended states.

Finally, assuming that particles in extended states make the largest
contribution to conductivity, we calculate the resistivity for different
concentrations and magnetic fields and compare with experiment.

We conclude that for the model can be used successfully to represent the
transport properties of the systems under consideration.
\end{abstract}


%\pagebreak

\section{INTRODUCTION}

The discovery of ''colossal ''magnetoresistance in La$_{1-x}$Sr$_x$MnO$_3$
type compounds \cite{helm} and its relation to possible applications to
magnetoresistance (MR) devices has attracted the attention of the physics
community in the last times.

Before the discovery of ''colossal '' MR, the earlier studies by Jonker and
Van Santen \cite{jova} established a temperature-doping phase diagram
separating metallic ferromagnetic from insulating antiferromagnetic phases.
Zener \cite{zener} proposed a ''Double Exchange '' (DE) mechanism to
understand the phase diagram of these compounds and the intimate link
between their magnetic and transport properties. This DE mechanism was used
by Anderson and Hasegawa \cite{andha} to calculate the ferromagnetic
interaction between two magnetic ions, and by de Gennes \cite{degen} to
propose canting states for the weakly doped compounds. Kubo and Ohata \cite
{kuo} used a spin wave approach to study the temperature dependence of the
resistivity at temperatures well below the critical temperature and a mean
field approximation at T near T$_c.$ . Mazzaferro, Balseiro and Alascio \cite
{jorge} used a mixed valence approach similar to that devised for TmSe
combining DE with the effect of doping to propose the possibility of a metal
insulator transition in these compounds.

More recently, a wealth of experimental results have been obtained on the
transport, optical, spectroscopic and thermal properties of these materials
under the effects of external magnetic fields and pressures \cite{todos}.

From the theoretical point of view, Furukawa \cite{furu} has shown that DE
is essential to the theory of these phenomena, while Millis {\it et al}. 
\cite{millis} have argued that DE alone is not sufficient to describe the
properties of some of the alloys under consideration and have proposed that
polaronic effects play an important role. In a previous paper we have
explored a semi-phenomenological model that includes the effect of disorder
in the transport properties \cite{alal}. M\"{u}ller-Hartmann and Hirsch \cite
{muhi} have pointed out that a new phase appears in the proper derivation of
the effective hopping, but have not studied its effect in the physical
properties of the systems under consideration.

Quantitative comparison between calculated and measured resistivities is
scarce, with some exceptions as is the case in \cite{exp1} where the
connection between magnetization and resistivity is clearly shown.

In our previous paper \cite{alal} we treat the Hamiltonian proposed for
these systems using an alloy analogy approximation to the exchange terms and
including the effects of disorder by introducing a continuous distribution
of the diagonal site energies.

Here we continue that treatment by proposing a Free Energy that allows to
determine the magnetization as a function of temperature. We then proceed to
find the Fermi energy and the mobility edge (ME) as functions of
temperature. Finally, assuming that the conductivity is dominated by
particles occupying extended states, we draw resistivity-vs-temperature
curves. We compare our results with experiments in single crystals of La$%
_{1-x}$Sr$_x$MnO$_3$ reported by Tokura {\it et al}.\cite{exp1} for four
different values of the concentration of Sr and four different values of the
magnetic field finding that the model allows a clear description of the
experimental results.

In Section II we describe the results of our previous paper and the
approximations made to obtain the conductivity, and Section III is devoted
to comparison with experiment and discussion of the results.

\section{MODEL}

In our previous paper\cite{alal} we consider a simplified model Hamiltonian
given by 
\[
H_m=\sum_{i,\mu }\epsilon _ic_{i\mu }^{\dagger }c_{i\mu }-t\sum_{<i,j>,\mu
}c_{i\mu }^{\dagger }c_{j\mu } 
\]
\begin{equation}
\hspace{10mm}+U\sum_ic_{i\uparrow }^{\dagger }c_{i\uparrow }c_{i\downarrow
}^{\dagger }c_{i\downarrow }-J\sum_i\vec{S}.\vec{\sigma}\,,
\end{equation}
where $c_{i\mu }^{\dagger }$, $c_{i\mu }$ creates and destroys an itinerant
electron with spin $\mu $ at site $i$ respectively, $\epsilon _i$ are the
site diagonal energies that depend on the site neighborhood. For simplicity
we consider only one E$_g$ orbital per site. If we were to include the two
degenerate orbitals, we would need to consider also the Coulomb and exchange
interactions between them to produce the Mott insulating states at both ends
of the concentration range. $\vec{S}$ and $\vec{\sigma}$ are the Pauli
matrices for spin $\frac 12$ at site $i$ for localized and itinerant
electrons respectively. $\epsilon _i$ is the on-site energy, $t$ the hopping
parameter between nearest neighbors, $U$ the on-site Coulomb repulsion
between two itinerant electrons, and $J$ is the ferromagnetic ($J>0$)
coupling between the localized and itinerant electrons. This Hamiltonian
represents each Mn$^{4+}$ ion at site $i$ by a spin $S_i=\frac 12$, on which
one electron can be added to produce Mn$^{3+}$ . When an electron is added
in the d-shell of site $i$, an exchange coupling $J$ is included to favor
parallel alignment of the added electron to the already existing spin \cite
{kasu}. Also to avoid the possibility of Mn$^{2+}$ we include a strong
Coulomb repulsion $U$ and we take $U\to \infty $.Without losing essential
physics we simplify further by taking only the $z$ component of the exchange
interaction. Thus the states of the system are characterized by itinerant
electrons moving on a frozen distribution of localized up or down spins. To
obtain site Green functions and thus local density of states for this
problem, we ignore at the start the site dependence of the diagonal
energies: i.e. we set $\epsilon _i=\epsilon $ and we use an alloy analogy
approximation to obtain the effect of $J$ (assumed larger than $t$) in the
electronic band structure of the model. Using the Renormalized Perturbation
Expansion \cite{eco} in the manner described in \cite{alal} we obtain the
corresponding local Green functions and the average density of states for
spin up and down. The densities of states for each spin split into two bands
centered at $E_{\pm }=(\epsilon \pm J)$ with weights and widths that depend
on the number of sites with each spin $S_z=+1/2$ or $-1/2$. i.e. they depend
on the magnetization of the system. The electronic structure of the
compounds consists of essentially four bands, two for spin up and two for
spin down, The splitting between the up and down bands is given by the
intra-atomic exchange energy $J$, their weight and width by the normalized
magnetization $m=2<S>$. The Fermi level falls always in the lower bands so
that the transport properties are determined by these bands. Consequently,
for $J>>\sqrt{K}t$ , where $K$ is the connectivity, using the site density
of state (Eq. (11) in Ref.\cite{alal}) the averaged density of states per
site reduces to 
\begin{equation}
\rho _{0\mu }(\omega )=\frac{\nu _\mu (K+1)\sqrt{4Kt^2\nu _\mu -(\omega -E)^2%
}}{2\pi |(K+1)^2t^2\nu _\mu -(\omega -E)^2|}\,.
\end{equation}
where $E=(\epsilon -J)$ and $\nu _\mu =(1+\mu $ $m)/2$ ($\mu =\pm $ for up
and down spin respectively).

At this point, we introduce the effect of the disorder originated by the
substitution of some of the rare earth ions by Sr,Ba or Ca. We assume that
this can be described within the model by making the diagonal energies site
dependent. As is well known, since Anderson's original paper \cite{andy} a
distribution of diagonal energies produces localization of the electronic
states from the edges of the bands to an energy within them which is called
''mobility edge'' (ME). The precise position of the ME is difficult to
calculate and different localization criteria result in different values for
it \cite{lie}. However, we do not aim here to an absolute value for the ME
but rather to its change with respect to the Fermi level when the
magnetization changes from saturation to cero. For this reason we assume
that there is no localization before disorder and for simplicity,we use a
Lorentzian distribution of energies \cite{lloyd} ( width $\Gamma $ ) and the
Ziman criterium of localization\cite{zim}.

From the ensemble-averaged Green function we obtain densities of states. 
\begin{equation}
\rho _\mu (\epsilon )=\int_{-\infty }^{+\infty }\rho _{0\mu }(\epsilon
^{\prime })L(\epsilon -\epsilon ^{\prime })d\epsilon ^{\prime }\,,
\end{equation}
where $L(x)$ is a Lorentz distribution given by 
\begin{equation}
L(x)=\frac \Gamma {[\pi (x^2+\Gamma ^2)]}\,.
\end{equation}

Within this {\it comparative} approach one can make the further
approximation of replacing in Eq.(3) $\rho _{0\mu }$ by a square density of
states with the same width $W_\mu =2t\sqrt{K\text{ }\nu _\mu }$ and the same
weight $\nu _\mu $ to obtain,

\begin{equation}
\rho _\mu (\epsilon )=\frac{\nu _\mu }{2\pi W_\mu }\left\{ \arctan \left[
(W_\mu -\epsilon )/\Gamma \right] +\arctan \left[ (W_\mu +\epsilon )/\Gamma
\right] \right\} ,
\end{equation}

which allows for analytical expressions for the number of particles $n,$ and
the internal energy $E$ as functions of the magnetization $m$ , and the
Fermi energy $\epsilon _F$. In some instances, when the Fermi level falls
too near the band edge, this approximation can differ from the more
realistic case where the density of states increases as $\sqrt{\epsilon }$ .
We will see bellow that this is the case for $n$=0.15 in the samples we use
to compare our results with.

To proceed further, we need an expression for the entropy of these system.
Again for $comparative$ purposes, we resort to the simplest possible form
compatible with our earlier approximations, that of a spin one half array of
sites:

\begin{equation}
S=\ln \left( 2\right) -\nu _{+}\ln \left( 2\nu _{+}\right) -\nu _{-}\ln
\left( 2\nu _{-}\right) .
\end{equation}

More accurate forms of the entropy valid in the mixed valence regime can be
used, see for example \cite{aligia}.

In the presence of a magnetic field $H$, the free energy per site is then ,

\begin{equation}
G=E-TS-\mu _BmH,
\end{equation}

where $T$ is the temperature and $\mu _B$ is the magnetic moment per site.

We proceed as follows: for each $n,$ we use (assuming $k_BT<<W_\sigma $)

\begin{equation}
n=\sum_\mu \int_{-\infty }^{\epsilon _F}\rho _\mu (\epsilon )d\epsilon \,,
\end{equation}

to obtain a relation between $n,m$ and $\epsilon _F$ from which $\epsilon _F$
can be determined numerically.

The free energy is then a function of $m$ and $T$ only and allows, by
minimization, to determine $m(T)$. The resulting $m(T)$ (Shown in Fig.1)
does not differ essentially from the law of corresponding states for spin
1/2. Having obtained $m(T)$ for each value of the parameters we can
determine the up and down mobility edges ( $B_{+}$ and $B_{-\text{ }}$) and
the Fermi Energy. They are also plotted as functions of temperature in Fig.1.

Following Mott and Davies \cite{moda} we calculate the transport properties
assuming that two forms of d.c. conduction are possible: thermally activated
hopping and excitation to the mobility edge. When the difference between the
Fermi level and the mobility edge $\Delta $ is not too large as compared to$%
\ k_BT$ , the conductivity is dominated by particles in the extended states,
and is given by the usual relaxation time form,

\begin{equation}
\sigma =\frac{e^2}{3a^3}*\sum_\mu \left\{ \int_{-\infty }^\infty v_\mu
^2(\epsilon )\tau _\mu (\epsilon )\rho _\mu (\epsilon )(-\frac{\partial
f(\epsilon )}{\partial \epsilon })d\epsilon \,\right\} ,
\end{equation}

in which $a$ is the Mn-Mn distance in the simple cubic lattice, $f(\epsilon
) $ is the Fermi function. We assume that the relaxation time $\tau _\mu $
is a step function equal to zero for $\epsilon <B_{\mu \text{ }}$ and takes
a value $\tau _o$ related to the minimum metallic conductivity for $\epsilon
>B_{\mu \text{ }},$where according to Ref. \cite{lloyd} $B_\mu =-\sqrt{%
t^2K^2 \text{ }\nu _\mu -\Gamma ^2}$ . Further replace $v_\mu (\epsilon )$
by its average $v_\mu ^2=W_\mu ^2a^2/2\hbar ^2$ to obtain: 
\begin{equation}
\sigma =\frac{e^2\tau _o}{6\hbar ^2a}*\sum_\mu \left\{ W_\mu ^2*\int_{B_\mu
}^\infty \rho _\mu (\epsilon )\,(-\frac{\partial f(\epsilon )}{\partial
\epsilon })d\epsilon \right\} .
\end{equation}

An Anderson transition takes place when $B_\mu $ vanishes. For $(t^2K^2$ $%
\nu _\mu -\Gamma ^2)<0$ all eigenstates became localized.

\section{RESULTS AND DISCUSSION}

In what follows we take $K=5$ appropriate to describe the simple cubic
lattice of the Mn sites and $t=1$ fixes the scale of energies. As a
consequence of the structure of the model and of the approximations that led
us to this point, the model becomes symmetric under electron -hole
transformation in the lower spin up and down bands.

For $n=0.5$ , the Fermi energy vanishes independently of the value of the
magnetization and one can obtain an analytical expression for the free
energy, from which we derive $T_C:$ 
\begin{equation}
T_C=[(\Gamma ^2+30t^2)\arctan (\sqrt{10}t/\Gamma )-\Gamma \sqrt{10}t]/(8\pi 
\sqrt{10}t).
\end{equation}

Connected to the transport properties we can define a characteristic
temperature $T_M$ at which the mobility edge crosses the Fermi level. Notice
however that this crossing does not imply any discontinuous change in the
resistivity, the only non-analyticity occurs at $Tc$. For $\ n=0.5$ we
obtain an explicit expression for $T_M:$ 
\begin{equation}
T_M=\frac 1{4\pi \ln (\frac{1+m_c}{1-m_c})}\sum_\mu \mu [(3A_\mu ^2-\Gamma
^2)\arctan (A_\mu /\Gamma )-2A_\mu \Gamma \ln (A_\mu ^2+\Gamma ^2)],
\end{equation}

where $A_\mu =2t\sqrt{Km_\mu }$ , $m_\mu =(1+\mu $ $m_c)/2$ , and $m_c=\frac{%
2\Gamma ^2}{t^2K^2}-1$. In Fig. 2 we show $T_C$ and $T_M$ as a function of $%
\Gamma $ for $\ n=0.5$.

In what follows we consider $n<0.5$ and identify $n$ with the number of
holes, which we take to be equal to the concentration of divalent component
of the alloy. We define as insulator the state where the Fermi level falls
below the ME ( $\Delta =(B_{+}-\epsilon _F)>0$) . So that, for small $\Gamma 
$ the Fermi level falls above the ME ( $\Delta <0$) and only the metallic
state appears. When $\Gamma $ increases, $\Delta $ reduces and, finally $%
\Delta =0$ for a critical value $\Gamma _{-}=\sqrt{0.5K^2t^2-\epsilon _F^2}$
(where $m_c=0$ and $T_C=T_M$). When $\Gamma $ is increased from $\Gamma _{-}$
, $T_M$ reduces and finally $T_M=0$ at a critical value $\Gamma _{+}=\sqrt{%
K^2t^2-\epsilon _F^2}$ . Above $\Gamma _{+}$ the system remains insulating
at all temperatures. Consequently, only for $\Gamma _{-}<\Gamma <\Gamma _{+ 
\text{ }}$the transition between metallic and insulating regimes appears.
All these facts are depicted in Fig. 2 for $n=0.5$ ($\epsilon _F=0$ ). Note
the similarity of $Tc$ vs $\Gamma $ with $Tc$ versus electron-phonon
coupling in Millis {\it et al}.\cite{millis}

In Fig. 3 we show $T_C$ and $T_M$ as functions of $\ n$ for some values of $%
\Gamma $. As a consequence of the density of states being modified by
disorder, the Curie temperatures decrease with $\Gamma $, while the increase
with $n$ is just a consequence of the energetics of the bands. Tentative
fitting of the calculated resistivity with the data on La$_{1-x}$Sr$_x$O$_3$
of reference \cite{exp1} gives a value of 1.8 t for $\Gamma .$

In Fig. 4 we have tried tentatively to fit the logarithm of the resistivity
as obtained from Eq.(10) to the measurements of Tokura {\it et al}. \cite
{exp1}.We have chosen to compare with these samples to avoid the
complications that arise from strong coupling to the lattice in the smaller
radius compounds \cite{hwan}. To do that, we fix arbitrarily the value of $%
\Gamma $ at 1.8 $t$. We let$\ t$ vary from sample to sample to fit $\ Tc.$
Starting with the curve corresponding to $x$ or $n$=0.175 we choose t=1704 K
and change to t=1529 K for n= 0.2, to t=1216 K for n=0.3, and to t=1600 K
for n=0.15. These values of t correspond to bandwidths that range between
1.3 eV to 0.93 eV. We then multiply the values of each calculated
resistivity by a constant ( in the logarithmic plot corresponds to shifting
the curves up and down ) to fit approximately the value at the maximum. This
last constant corresponds to different values of $\tau _o$ in Eq.10. which
range in the 10$^{-15}$ to 10$^{-16}$ sec. These $\tau _o$'s correspond to
the minimum conductivity defined in Mott and Davis \cite{moda}. We can see
that the fitting is better in the more ''metallic '' samples than in the $n$%
=0.15 sample where one could expect the contribution of localized states to
be larger and the model results differ more from experiment. Indeed, as
pointed out above, the resistivity calculated with the square density of
states differs even more from experiment that the one shown in the Fig.4,
which is calculated with the more realistic density of states of Eq. 2 .

We conclude from the comparison that the model allows to characterize the
resistivity behavior of different samples by two parameters, one associated
to the degree of disorder ($\Gamma $), and the other to the hopping energy $%
t $. The values of the hopping energy $t$ can be affected by displacement of
the oxygen atoms, or by polaronic or other many body effects.

In Fig. 5 we show the magnetic field effect on the resistivity and compare
again with the results obtained in \cite{exp1}. Here again, we take $\Gamma
=1.8,$ $t=1789$ $K$ and select $\tau _o=0.96*10^{-14}$ to fit the $H=8$ T
curve. We take $\mu _B=0.964*10^{-20}$ erg/Gauss to fit the rest of the
curves. Indeed, the fitting of resistivity curves in the absence of magnetic
field should be taken with care because of the effect of magnetic domains
walls.

Three main interactions should be incorporated in a more complete
description of the whole family of ''colossal magnetoresistance '' Mn
perovkites.

1. Static and dynamic lattice effects can modify not only the values of both
parameters, $t$ and $\Gamma .$ but also the thermodynamics of the
transition, leading to first order transitions as those found in many of the
compounds\cite{kuwa}. The connection to the dynamics of the lattice has been
recently very elegantly demonstrated by Zhao {\it et al.}\cite{zhao}

2. Coulomb interactions between ions, that in combination with point 1 above
could also produce charge ordering and lead to the reentrant behavior found
in \cite{kuwa}.

3.Superexchange interactions between the localized spins, that lead to
canted states, as those found in Electron doped Ca$_{1-x}$Y$_x$Mn0$_3$ \cite
{tovar}.

The thermopower can also be calculated in a similar way. We will report
results for this quantity in a forthcoming paper. Measurements of this
quantity and resistivity in the same crystalline samples would be highly
desirable.

To summarize, we have shown that a very simple estimation of the effect of
disorder on the double exchange mechanism allows to understand resistivity
and magnetoresistivity of Sr doped La manganites. The most natural source of
disorder is the substitution of rare earth by Sr, Ca or Ba, but polaronic or
other many body effects may act in a similar way.

.\vspace{7mm}

{\large {\bf FIGURE CAPTIONS }} \vspace{3mm}

%\QTP{SçSçTSç}
Figure 1. Zero field magnetization (upper panel) and $\Delta
=(B_{+}-\epsilon _F)$ (lower panel) are plotted as a function of the
normalized temperature ($T/T_C$) for $\Gamma =1.8$, $K=5$, $t=1$, and
different values of doping $n$.

%\QTP{SçSçTSç}
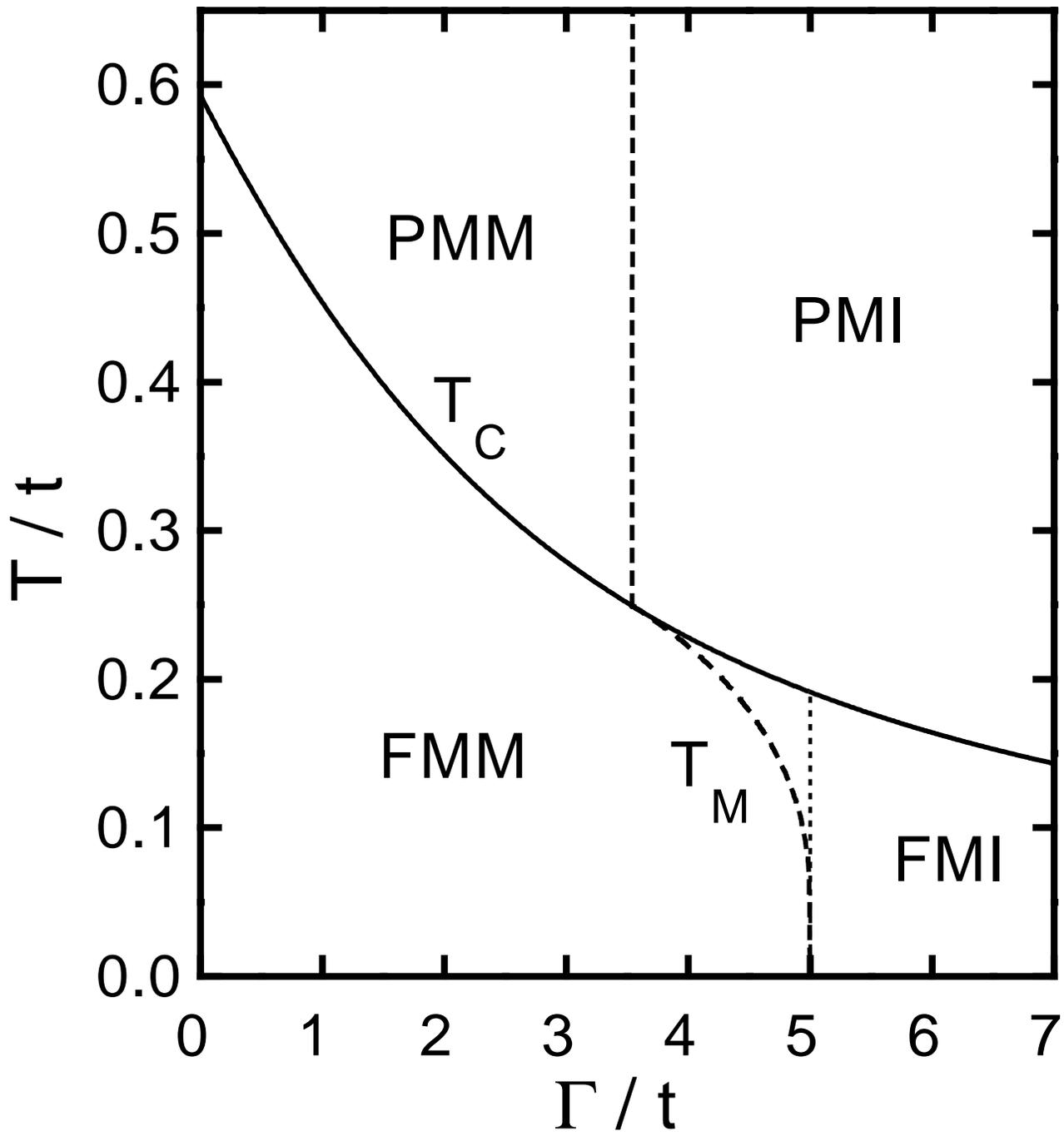
Figure 2. Phase diagram for $n=0.5$ ($\epsilon _F=0$ ). Ferromagnetic $T_C$
(solid line) and metal-to-insulator $T_M$ (dashed line) transition
temperatures vs $\Gamma $, for $H=0$, $K=5$, and $t=1.$ Regions labelled as
FMM (ferromagnetic metal: $m\neq 0$ and $\Delta <0$), FMI (ferromagnetic
insulator: $m\neq 0$ and $\Delta >0$), PMM (paramagnetic metal: $m=0$ and $%
\Delta <0$), and PMI (paramagnetic insulator: $m=0$ and $\Delta >0$). Dotted
line is a guide to the eye.

%\QTP{SçSçTSç}
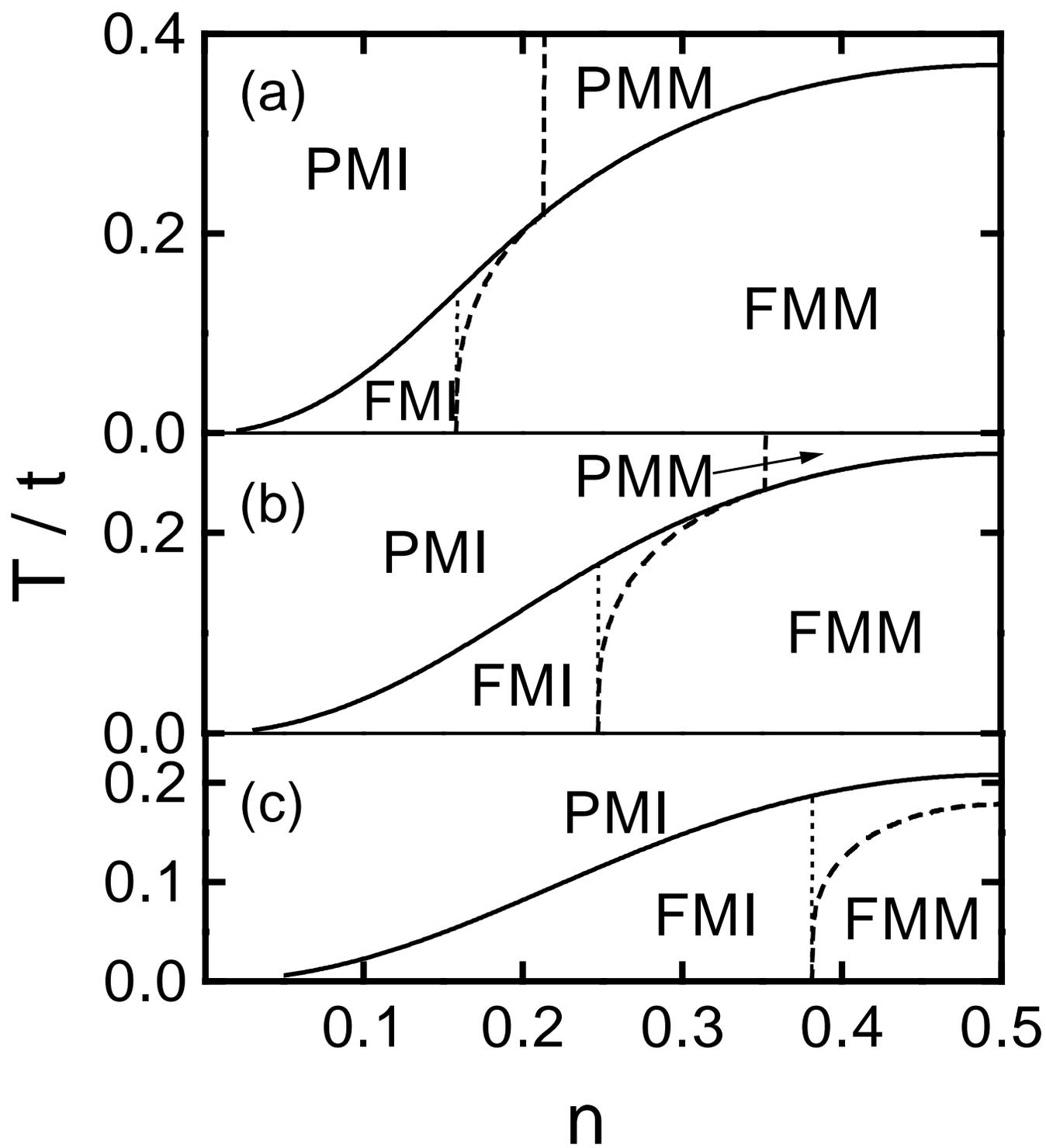
Figure 3. Phase diagram. Ferromagnetic $T_C$ (solid lines) and
metal-to-insulator $T_M$ (dashed lines) transition temperatures vs doping $n 
$, for $H=0$, $K=5$, $t=1$, and different values of $\Gamma $: (a) $\Gamma
=1.8$, (b) $\Gamma =3$, and (c) $\Gamma =4.5$. Regions labelled as FMM
(ferromagnetic metal: $m\neq 0$ and $\Delta <0$), FMI (ferromagnetic
insulator: $m\neq 0$ and $\Delta >0$), PMM (paramagnetic metal: $m=0$ and $%
\Delta <0$), and PMI (paramagnetic insulator: $m=0$ and $\Delta >0$). Dotted
lines are a guide to the eye.

%\QTP{SçSçTSç}
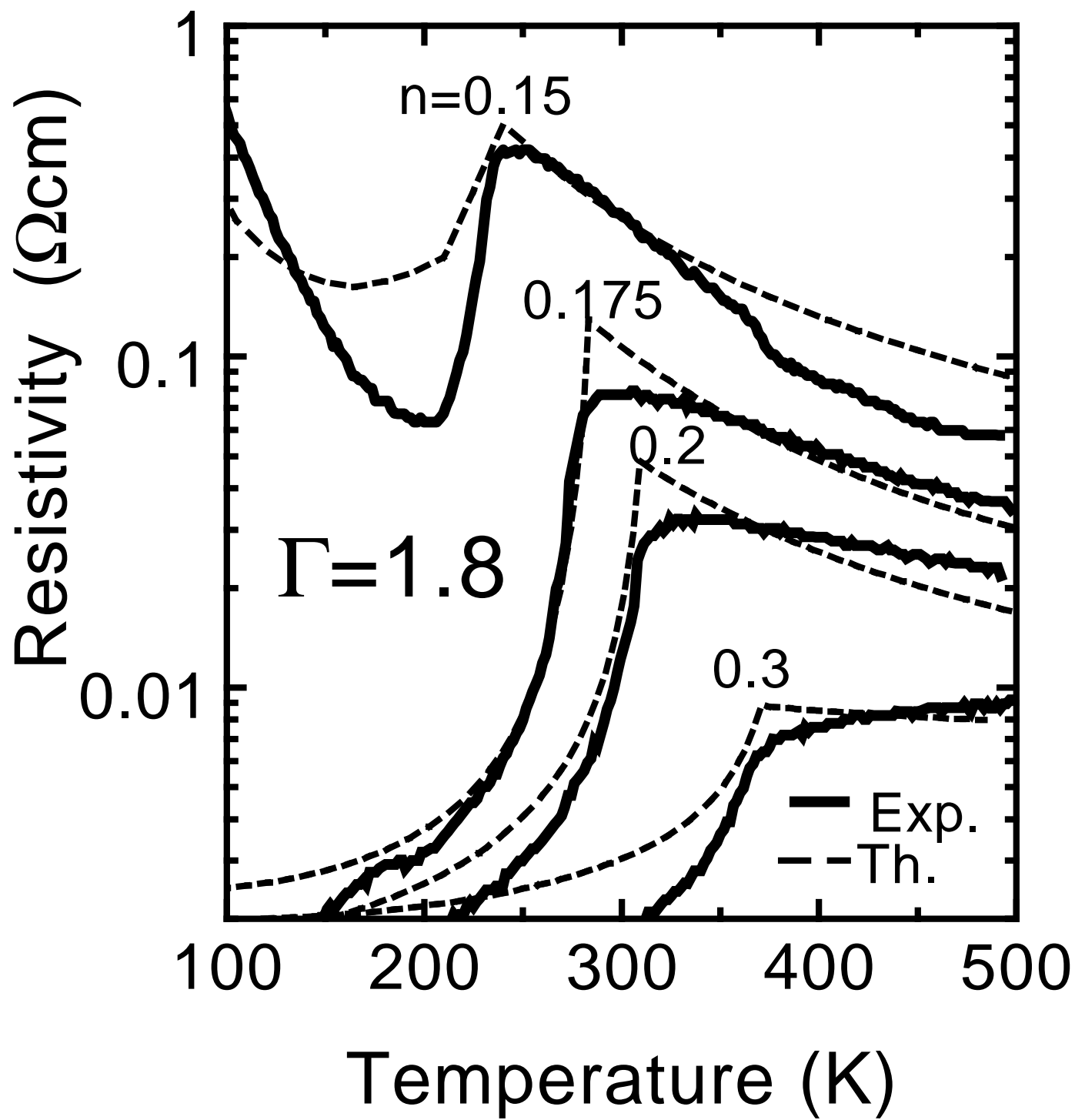
Figure 4. Zero field resistivity (solid lines) on a logarithmic scale vs
temperature in La$_{1-n}$ Sr$_{n\text{ }}$Mn O$_{3\text{ }}$taken from Ref. 
\cite{exp1}. The dashed lines represent the fits with Eq. (10) for $t=1$ , $%
\Gamma =1.8$, $K=5$, and the corresponding values of doping: $n=0.15$ , $%
n=0.175$, $n=0.2$, and $n=0.3$.

%\QTP{SçSçTSç}
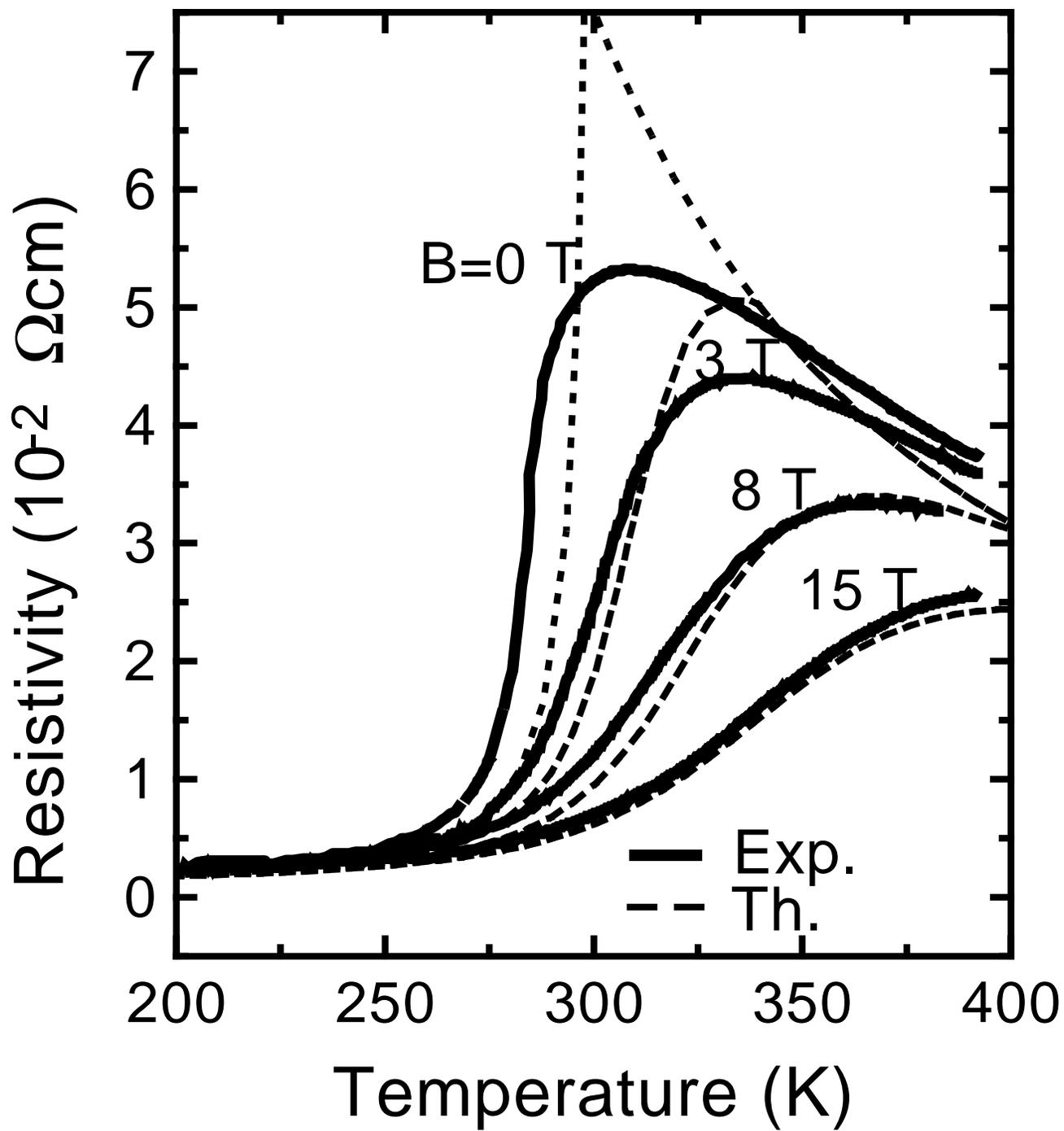
Figure 5. Magnetoresistance in La$_{1-n}$ Sr$_{n\text{ }}$Mn O$_{3\text{ }}$(%
$n=0.175$). The solid lines show the temperature dependence of resistivity
in magnetic fields taken from Ref. \cite{exp1}. The dashed lines represent
the fits with Eq. (10) for $t=1$ , $\Gamma =1.75$, $K=5$, and the
corresponding values of magnetic field: $B=0$ , $B=3$ T, $B=8$ T, and $B=15$
T.


\begin{references}
\bibitem{helm}  R. von Helmholt, J. Wecker, B. Holzapfel, L. Schultz, and K.
Samwer, Phys. Rev. Lett. {\bf 71}, 2331 (1993).

\bibitem{jova}  G. H. Jonker and J. H. van Santen, Physica {\bf 16}, 337
(1950); J. H. van Santen and G. H. Jonker, Physica {\bf 16}, 599 (1950).

\bibitem{zener}  C. Zener, Phys. Rev. {\bf 82}, 403 (1951).

\bibitem{andha}  P. W. Anderson and H. Hasegawa, Phys. Rev. {\bf 100}, 675
(1955).

\bibitem{degen}  P. G. de Gennes, Phys. Rev. {\bf 118}, 141 (1960).

\bibitem{kuo}  K. Kubo and N. Ohata, J. Phys. Soc. Jpn. {\bf 33}, 21 (1972).

\bibitem{jorge}  J. Mazzaferro, C. A. Balseiro, and B. Alascio, J. Phys.
Chem. Solids {\bf 46}, 1339 (1985).

\bibitem{todos}  Y. Moritomo, A. Asamitsu, and Y. Tokura, Phys. Rev. B {\bf %
51}, 16491 (1995); Y. Okimoto, T. Katsufuji, T. Ishikawa, A. Urushibara, T.
Arima, and Y. Tokura, Phys. Rev. Lett. {\bf 75}, 109 (1995); S. W. Cheong,
H. Y. Hwang, P. G. Radaelli, D. E. Cox, M. Marezio, B. Batlogg, P. Schiffer,
and A. P. Ramirez, Proceedings of the ''Physical Phenomena at High Magnetic
Fields - II'' Conference, Tallahassee, Florida. World Scientific, to be
published;; M. C. Martin, G. Shirane, Y. Endoh, K. Hirota, Y. Moritomo, and
Y. Tokura, To be published; R. Mahendiran, R. Mahesh, A. K. Raichaudhuri,
and C. N. R. Rao, Solid State Commun. {\bf 94}, 515 (1995); H. L. Ju, J.
Gopalakrishnan, J. L. Peng, Qi Li, G. C. Xiong, T. Venkatesan, and R. L.
G\`{\i}eene, Phys. Rev. B {\bf 51}, 6143 (1995); M. K. Gubkin, A. V.
Salesskii, V. G. Krivenko, T. M. Perekalina, T. A. Khimich, and V. A.
Chubarenko, JETP Lett. {\bf 60}, 57 (1994).

\bibitem{furu}  N. Furukawa, J. Phys. Soc. Jpn. {\bf 63}, 3214 (1994).

\bibitem{millis}  A. J. Millis, P. B. Littlewood, and B. I. Shrainman, Phys.
Rev. Lett. {\bf 74}, 5144 (1995).

\bibitem{alal}  R. Allub and B. Alascio, Solid State Commun. {\bf 99},
613 (1996).

\bibitem{muhi}  E. M\"{u}ller-Hartmann and J. E. Hirsch, preprint.

\bibitem{exp1}  Y. Tokura, A. Urushibara, Y. Moritomo, T. Arima, A.
Asamitsu, G. Kido, and N. Furukawa, J. Phys. Soc. Jpn. {\bf 63}, 3931 (1994).

\bibitem{kasu}  T. Kasuya, Prog. Theor. Phys. {\bf 16}, 45 (1956).

\bibitem{eco}  See, e.g., E. N. Economou, {\it Green's Functions in Quantum
Physics} Springer Series in Solid-State Sciences {\bf 7}, Ed. P. Fulde.

\bibitem{andy}  P. W. Anderson, Phys. Rev. {\bf 109}, 1492 (1958).

\bibitem{lie}  D. C. Licciardello and E. N. Economou, Phys. Rev. {\bf 11},
3697 (1975).

\bibitem{lloyd}  P. Lloyd, J. Phys. C {\bf 2}, 1717 (1969).

\bibitem{zim}  J. M. Ziman, J. Phys. C {\bf 2}, 1230 (1969).

\bibitem{aligia}  A. A. Aligia Thesis, Instituto Balseiro (1984).

\bibitem{moda}  N. F. Mott and E. A. Davis, {\it Electronic Processes in
Non-Crystallyne Materials}, Oxford University Press (1971).

\bibitem{hwan}  H. Y. Hwang, S.W. Cheong, P. G. Radelli, M. Marezio, and B.
Batlogg, Phys. Rev. Lett.{\bf 75}, 914 (1995).

\bibitem{kuwa}  H. Kuwahara, Y Tomioka, A. Asamitsu, Y. Moritomo, Y. Tokura,
Science {\bf 270}, 961 (1995).

\bibitem{zhao}  Guo-meng Zhao, K. Conder, H. Keller and K. A. Muller, Nature 
{\bf 381}, 676, (1996).

\bibitem{tovar}  J. Briatico, B. Alascio, R. Allub, A. Butera, A. Caneiro,
M. T. Causa, and M. Tovar. Czechoslovak J. Phys. {\bf 46, }S4 2013 (1996).
\end{references}
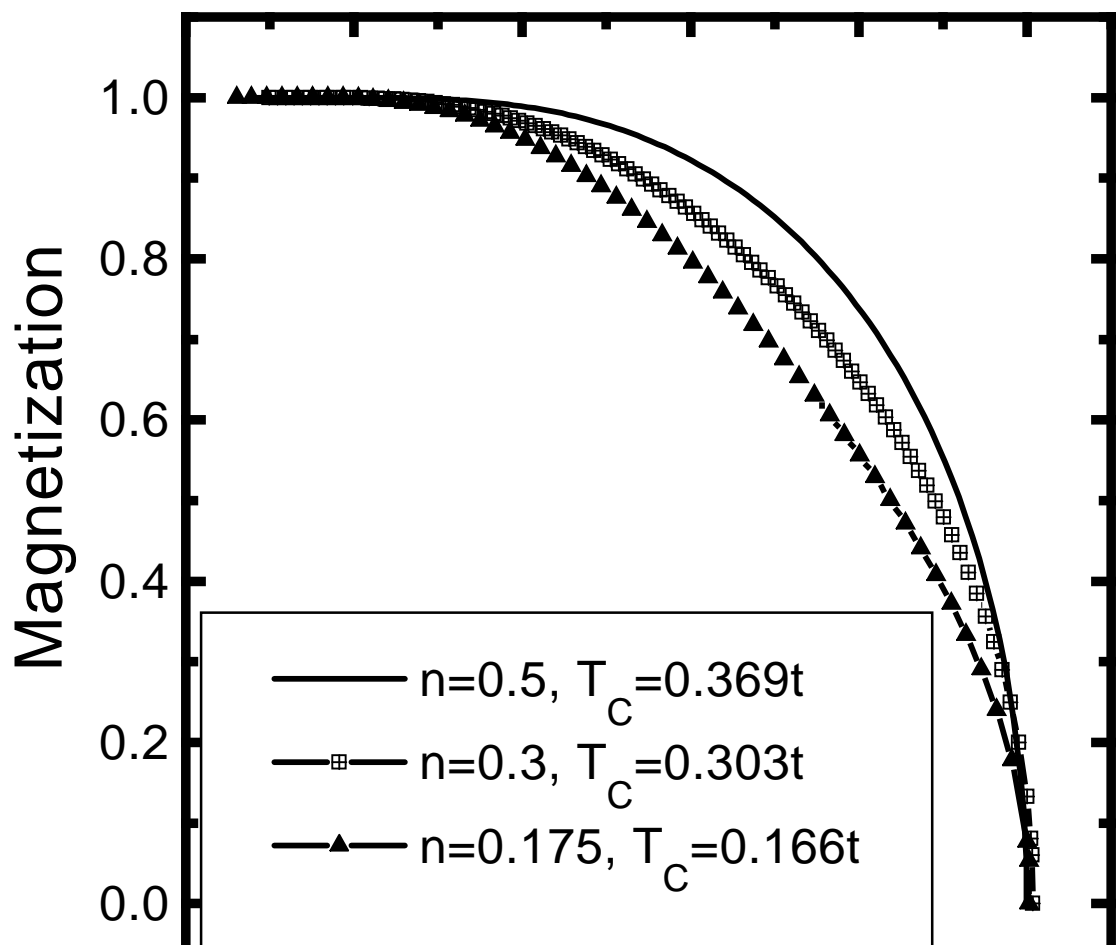
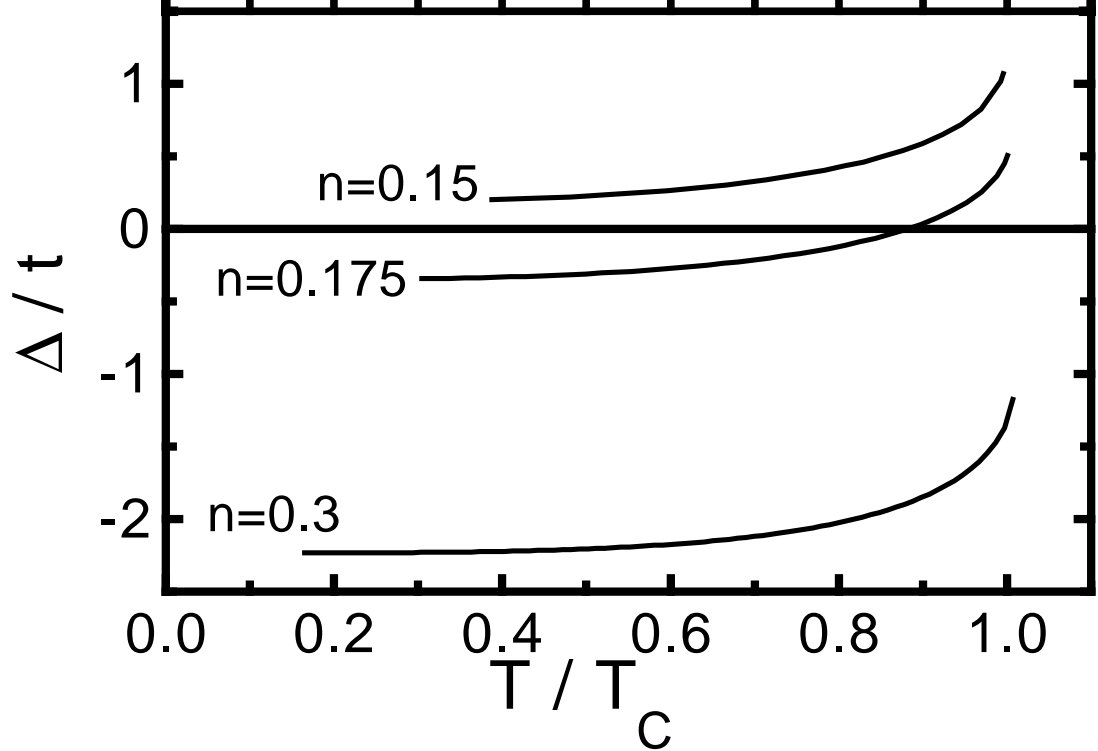
\end{document}